\documentclass[runningheads]{llncs}
\usepackage[T1]{fontenc}
\usepackage{xcolor}
\usepackage{graphicx}
\usepackage{multirow}
\usepackage{cite}
\usepackage{hyperref}
\usepackage{lscape}
\usepackage{dingbat}
\usepackage{color, colortbl}
\usepackage{algorithm}
\usepackage{algpseudocode}
\usepackage{amssymb}
\usepackage{graphicx}
\usepackage{bbm}
\begin{document}
%

\title{Optimizing Brain Tumor Segmentation with MedNeXt: BraTS 2024 SSA and Pediatrics}
%
%

\author{
    Sarim Hashmi \and
    Juan Lugo \and
    Abdelrahman Elsayed \and
    Dinesh Saggurthi \and
    Mohammed Elseiagy \and
    Alikhan Nurkamal \and
    Jaskaran Walia \and
    Fadillah Adamsyah Maani \and
    Mohammad Yaqub
}

\authorrunning{S. Hashmi et al.}

\institute{
    Mohamed bin Zayed University of Artificial Intelligence (MBZUAI)\\
    Abu Dhabi, United Arab Emirates\\
    \email{
        \{sarim.hashmi, juan.lugo, abdelrahman.elsayed, dinesh.saggurthi,\\
        mohammed.abdelaziz, alikhan.nurkamal, jaskaran.walia,\\
        fadillah.maani, mohammad.yaqub\}@mbzuai.ac.ae
    }\\
    \url{https://mbzuai.ac.ae}
}
\maketitle              
\begin{abstract}

Identifying key pathological features in brain MRIs is crucial for the long-term survival of glioma patients. However, manual segmentation is time-consuming, requiring expert intervention and is susceptible to human error. Therefore, significant research has been devoted to developing machine learning methods that can accurately segment tumors in 3D multimodal brain MRI scans. Despite their progress, state-of-the-art models are often limited by the data they are trained on, raising concerns about their reliability when applied to diverse populations that may introduce distribution shifts. Such shifts can stem from lower quality MRI technology (e.g., in sub-Saharan Africa) or variations in patient demographics (e.g., children). The BraTS-2024 challenge provides a platform to address these issues. This study presents our methodology for segmenting tumors in the BraTS-2024 SSA and Pediatric Tumors tasks using MedNeXt, comprehensive model ensembling, and thorough postprocessing. Our approach demonstrated strong performance on the unseen validation set, achieving an average Dice Similarity Coefficient (DSC) of 0.896 on the BraTS-2024 SSA dataset and an average DSC of 0.830 on the BraTS Pediatric Tumor dataset. Additionally, our method achieved an average Hausdorff Distance (HD95) of 14.682 on the BraTS-2024 SSA dataset and an average HD95 of 37.508 on the BraTS Pediatric dataset.Our GitHub repository can be accessed here: \href{https://github.com/python-arch/BioMbz-Optimizing-Brain-Tumor-Segmentation-with-MedNeXt-BraTS-2024-SSA-and-Pediatrics}{Project Repository}.

\keywords{BraTS \and Brain MRI \and Glioma \and Tumor segmentation \and MedNeXt  \and BraTS-SSA \and BraTS-PEDs } 

\end{abstract}

\section{Introduction}

Gliomas remain the most common and malignant type of primary brain tumor despite advances in understanding their pathophysiology \cite{gliomas}. Only about 20\% of individuals with glioma survive two years post-diagnosis, and they cause the highest number of cancer-related deaths in pediatrics. Their presence is visible through Magnetic Resonance Imaging (MRI) scans, which provide detailed information on the patient's internal structures, tissues, and organs. Correctly bounding the tumor and identifying its sub-regions from the MRI scan is a crucial first step in treatment options, including surgery, chemo, and radiation therapy.  Additionally, it is important for assessing treatment response and for the longitudinal monitoring of the patient\cite{owrangi2018mri}.

Radiologists rely on MRI scans to manually predict tumor classification and localization, which is time-consuming, labor-intensive, and susceptible to human error. These challenges are further compounded in low-income countries, where overburdened healthcare systems and a shortage of qualified specialists prevail \cite{bratsafrica}. As a result, there is a critical need for automated solutions, and machine learning pipelines for automatic tumor segmentation offer a promising tool to provide accurate and efficient measurements.

The Medical Image Computing and Computer Assisted Interventions (MICCAI) conference annually hosts various medical imaging competitions that draw research teams internationally. Among these is the BraTS challenge \cite{menze2014multimodal}, which consists of ten distinct brain tumor-related tasks this year. Originally, BraTS centered on adult glioma segmentation \cite{bakas2017advancing, bakas2017segmentation}, with datasets mainly sourced from the Global North. However, despite the advancements brought by this annual challenge, there have been concerns about the reliability of state-of-the-art models when applied to populations that introduce distribution shifts from the original dataset \cite{africadata}. Such shifts can result from lower quality MRI technologies (e.g., in sub-Saharan Africa) or differences in the population's anatomy (e.g., children's brains). Thus, the BraTS challenge expanded its dataset to include sub-Saharan African, pediatrics, and meningioma tumors. This paper focuses on the Africa and Pediatric tasks.

Since the 2014 challenge, deep learning has become the state-of-the-art for brain tumor segmentation, driven by rapid advancements in GPU technology and the availability of large datasets \cite{ferreira2024wonbrats2023adult}. Most solutions utilize the U-Net architecture, which features a contracting path similar to a convolutional network and an expansive path of up-sampling convolutions. Numerous enhancements have been made to the U-Net \cite{unet}, including the incorporation of residual connections, densely connected layers, and attention mechanisms \cite{ferreira2024wonbrats2023adult}.

This paper focuses on the MedNeXt architecture \cite{roy2023mednext}, a variation of U-Net that employs ConvNeXt \cite{liu2022convnet} blocks, for automatic brain tumor segmentation. We delve into the various preprocessing, ensemble techniques, and training modifications implemented to achieve high accuracy. The data used for performance assessment was obtained from standard clinical care for brain tumors and was meticulously annotated by radiologists and reviewed by neurologists to ensure accuracy \cite{bratsafrica, pedsdata}. The compounding improvements of these techniques result in a state-of-the-art machine learning pipeline for the BraTS Africa and Pediatric challenges.  

Our main contributions are the following:
\begin{itemize}
    \item A deep learning pipeline for Africa and Pediatric brain tumor segmentation for the corresponding BraTS 2024 tasks.
    \item An integration of the novel schedule free \cite{schedulefree}  optimizer into the training algorithm.
    \item A thorough analysis of finetuning and ensembling techniques, and their impact on performance. 
\end{itemize}

\section{Methods}

\subsection{Brain MRI Data}


Brain MRI is an important imaging technique in neurological diagnostics, offering high-resolution images of the brain's anatomy and function. A patient's MRI is divided into four key modalities, each detailing different aspects of medical interest. These are, T1-weighted imaging (T1) for detailed anatomical structure; T1 post galodinium contrast (T1Gd) for high vascularity pathologies; T2-weighted imaging (T2W) for detecting edema and inflammation; and T2 Fluid-Attenuated Inversion Recovery (T2-FLAIR) for enhanced lesion visibility \cite{menze2014multimodal}. Combining these modalities provides comprehensive insights for diagnosing and managing conditions such as strokes, tumors, multiple sclerosis, and traumatic brain injuries, ultimately improving patient outcomes.


The International Brain Tumor Segmentation (BraTS) challenge, ongoing since 2012, focuses on creating a benchmarking environment and dataset for delineating adult brain gliomas from MRI scans following the aforementioned modalities. In 2024, the challenge expands to approximately 4,500 cases, addressing diverse clinical and technical considerations. Key tasks include evaluating algorithms for post-treatment glioma, radiotherapy meningioma, brain metastasis, handling missing data, pathology analysis, and segmenting the tumor of specific demographics. In particular, this paper focuses on the challenges of BraTS-Africa and Pediatrics. 

\begin{figure}[!h]
    \centering
    \includegraphics[width=0.5\linewidth]{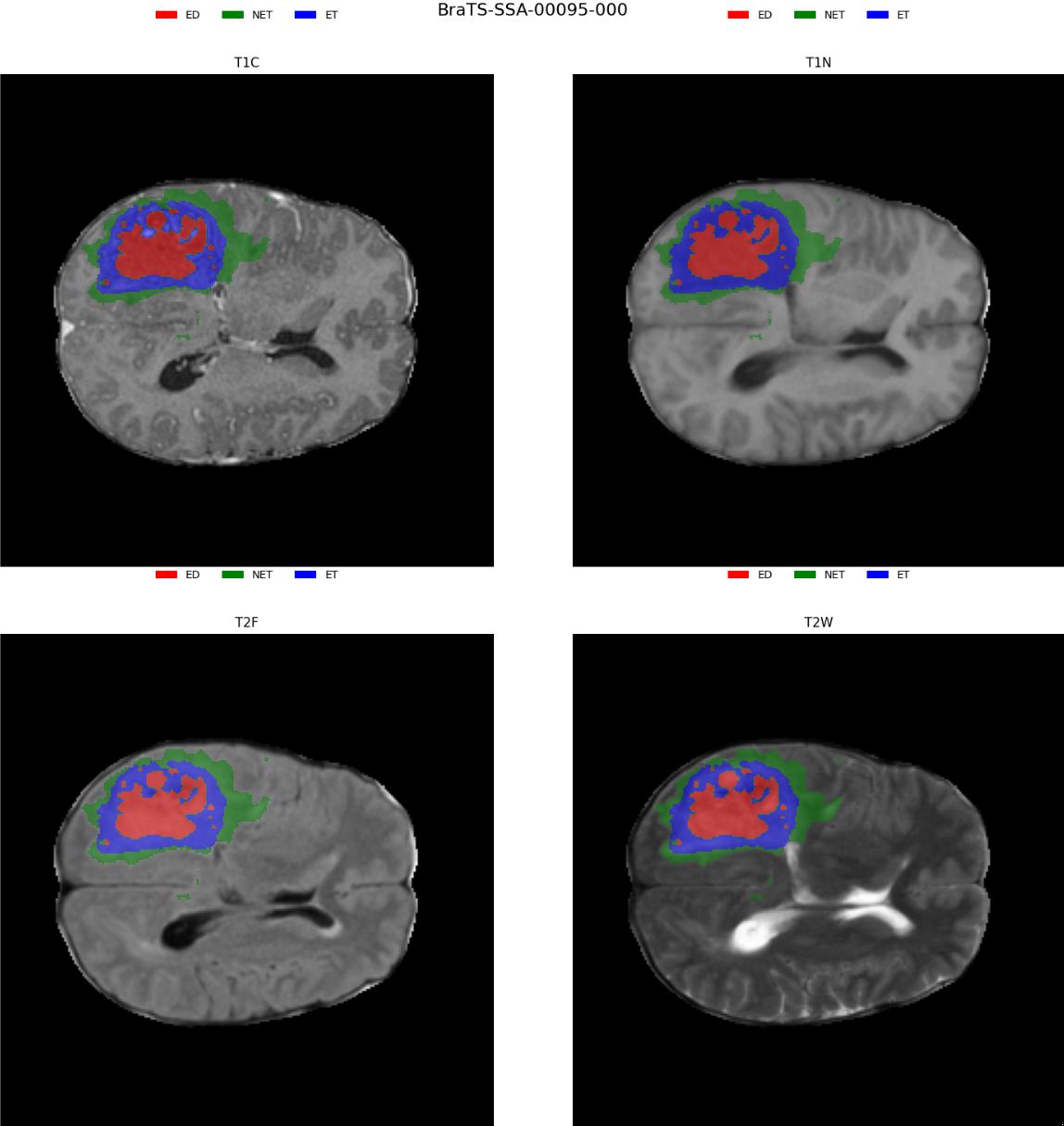}
    \caption{Cross sections of the four modalities obtained from a sample data-point from the provided dataset for the BraTS-Africa challenge along with the corresponding segmentation masks}
    \label{fig:ssa-fig}
\end{figure}

\noindent \textbf{BraTS-Africa dataset} The dataset \cite{bratsafrica} comprises multi-institutional structural MRI scans from 95 Sub-Saharan Africa patients. Out of these, 60 samples are used for training and 35 for validation. This year's task consists of identifying and segmenting the tumor into three sub-regions: the enhancing tumor (ET), the Non-enhancing tumor core (NETC), and the Surrounding non-enhancing FLAIR hyperintensity (SNFH). The segmentation labels in the dataset were annotated by board-certified radiologists and validated by expert neuroradiologists following BraTS protocols. To ensure the validity of the leaderboard, the ground truth for the 35 testing datapoints was not made public. Figure \ref{fig:ssa-fig} shows a sample of the training split. Additionally, BraTS SSA 2024 allows the use of the 1470 MRIs and labels of the BraTS-2023 Adult Glioma Challenge for further training \cite{baid2021rsna}. 

\begin{figure}[!h]
    \centering
    \includegraphics[width=\linewidth]{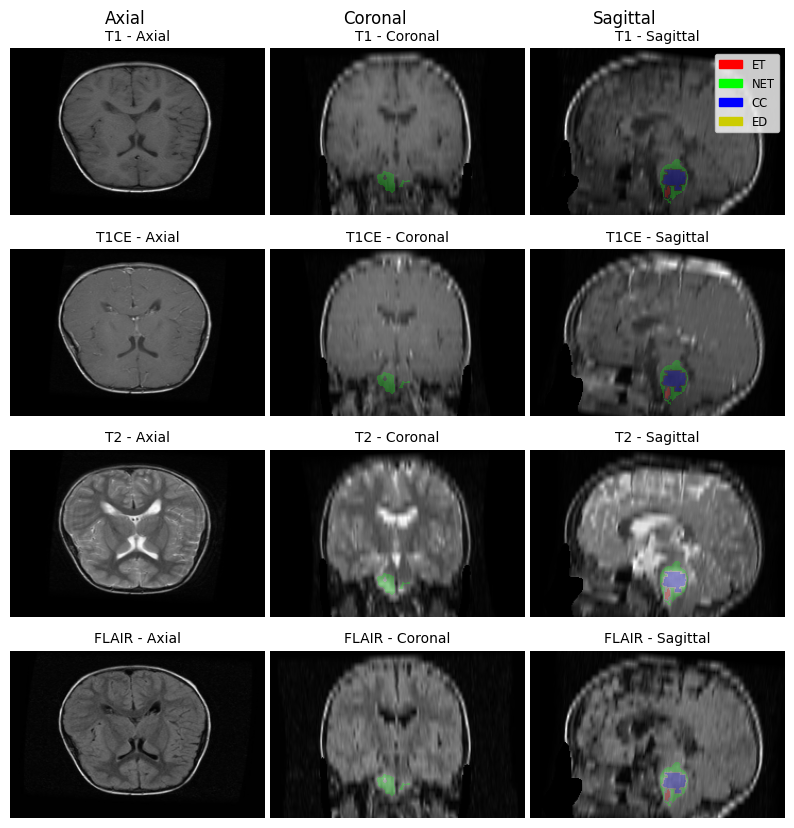}
    \caption{Different cross sections of the four modalities obtained from a sample data-point from the provided dataset for pediatrics challenge along with the corresponding segmentation masks}
    \label{fig:ped-fig}
\end{figure}

\noindent \textbf{BraTS PED}
The BraTS-PED dataset \cite{pedsdata} consists of multi-institutional conventional and structural MRI scans from 464 pediatric cases of high-grade glioma. There is heterogeneity in the images' quality due to the different protocols and equipment used. The data went through several preprocessing steps to ensure data integrity and patient anonymity. This year's task evaluates the accuracy of segmenting the tumor into six sub-regions of interest: the enhancing tumor (ET); the tumor core (TC); the whole tumor (WT); the non-enhancing tumor core (NETC); the cystic component (CC); and the peritumoral edema (ED). Out of these regions, only the first 3 are predicted by the model, and the latter 3 are inferred at testing time by the BraTS organizers. A sample from the dataset is shown in Figure \ref{fig:ped-fig}. A preliminary automated segmentation pipeline was used to obtain the segmentation labels, and the results were then manually refined by volunteer neuroradiologists. These refinements underwent iterative reviews and approvals by three board-certified neuroradiologists until the labels met the criteria for public release \cite{pedsdata}.


\noindent \textbf{Preprocessing} \label{sec:preprocessing} The challenge organizer has performed an initial preprocessing of the MRI scans by co-registering the four modalities to a standard SR124 template \cite{sr124}, isotropically interpolating to achieve a 1mm³ resolution. Skull-stripping was applied to the BraTS Africa Dataset, but not Pediatrics. All MRI images have a uniform size of 240 × 240 × 155. 

We further preprocessed the images cropping the foreground, normalizing voxels with non-zero intensities, and stacking the four modalities into a single image. We additionally crop the MRIs into patches of shape 128 x 160 x 112. We preprocessed all MRI scans in advance, stored them in ".npy" format, and loaded the Numpy arrays during training to speed up data loading and prevent the CPU bottleneck.



\subsection{MedNeXt}

\begin{figure}[!h]
    \centering
    \includegraphics[width=0.9\linewidth]{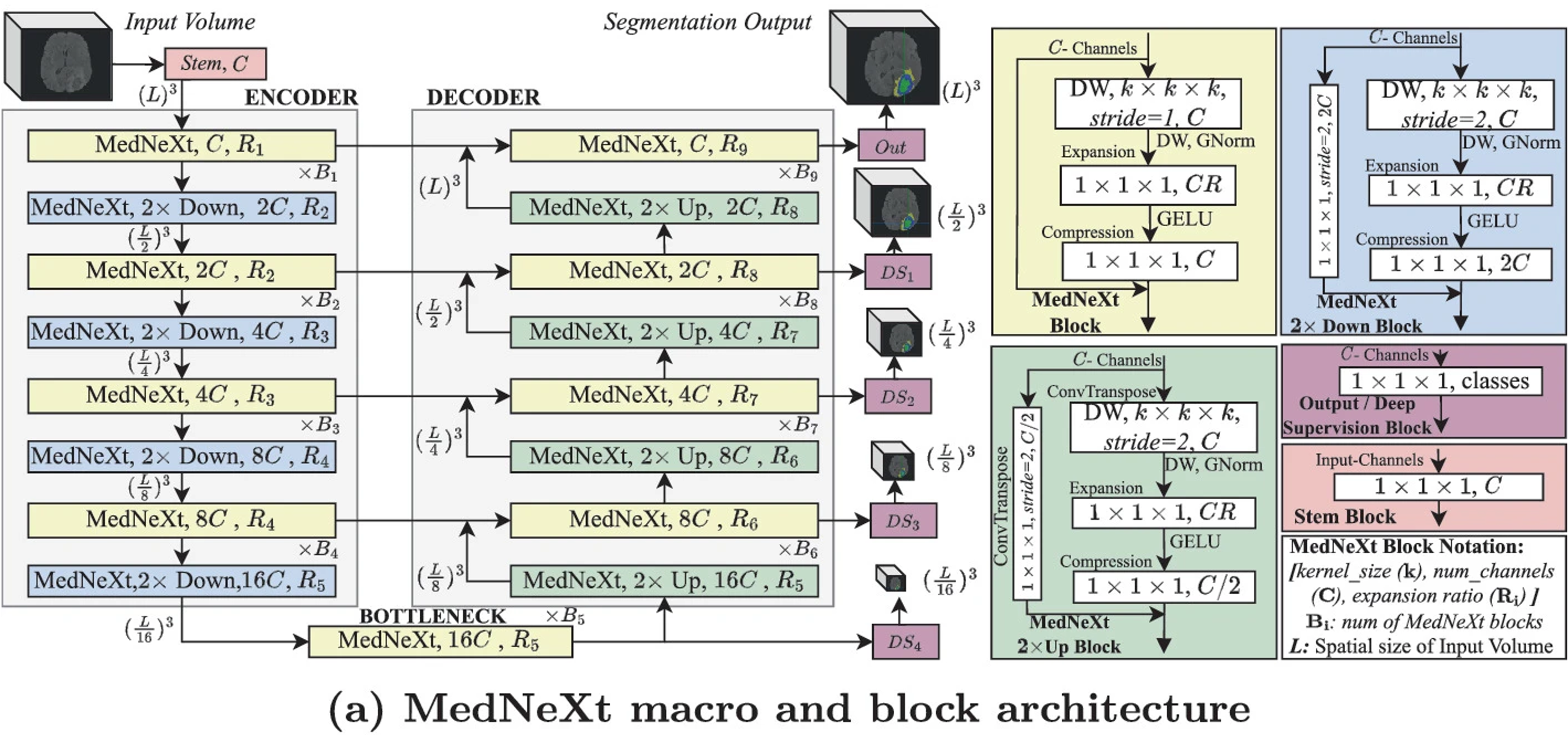}
    \caption{\textbf{(a)} Architectural design of the MedNeXt. The network has 4 Encoder and Decoder layers each, with a bottleneck layer. MedNeXt blocks are present in Up and Downsampling layers as well. Deep Supervision is used at each decoder layer, with lower loss weights at lower resolutions. All residuals are additive while convolutions are padded to retain tensor sizes. For further details, we refer to \cite{roy2023mednext}} 
    \label{fig:mednext-fig}
\end{figure}

Inspired by \cite{brats2,brats1} we adopted MedNeXt as our baseline model (Figure \ref{fig:mednext-fig}), following the standard implementation available on GitHub\footnote{https://github.com/MIC-DKFZ/MedNeXt}. MedNeXt is a state-of-the-art vision architecture that combines the long-range spatial representation of transformers with the inductive bias of ConvNets. It utilizes MedNeXt blocks arranged in a U-net like shape, with the characteristic encoder on the left side, bottleneck center, and decoder on the right side. Each "left-hand side" MedNeXt block contains three convolutions: first, a depth-wise convolution of dimension $k \times k \times k$ (($k \in \{3,5,7,9\}$) that emulates the large attention window of the vision transformer; second, an expansion layer, forming a transformer-like inverse bottleneck, where the number of channels is multiplied by an expansion ratio; and then a compression layer, where the number of channels is brought down again by $1 \times 1 \times 1$ kernels. Each "right-hand side" MedNeXt block follows the same structure but uses convolutions transposed to account for the required up-sampling. MedNeXt also includes deep supervision to prevent vanishing gradients and improve performance. We used the Base (B) and Medium (M) sizes for the architecture.



\subsection{Model Development}


\subsubsection{Model training}
We conducted our experiments on an NVIDIA GPU with 24 GB of memory. For the SSA challenge, we combined the BraTS Adult Glioma and BraTS Africa Training datasets. In both tasks, Pediatrics and Africa, the model’s input size was set to 128x160x112 voxels, following our preprocessing. Using 5-fold cross-validation on the training set, we trained the MedNeXt-B and MedNeXt-M models for the BraTS Africa dataset. Both models were implemented with a kernel size of 3 for 150 epochs. On the other hand, only MedNeXt-B was trained with a kernel size of 3 and 75 epochs for the BraTS pediatrics dataset. Optimization was performed using the Schedule-free AdamW optimizer \cite{schedulefree}, recently proposed by the Fundamental AI Research Team at Meta, with a learning rate of 0.0027 for BraTS Africa, learning rates of 0.0027 and 0.0005 for BraTS pediatrics, and no weight decay. We used a batch size of 2 for training. For the objective function (loss), deep supervision, brain MRI preprocessing, and data augmentation, we followed the approach outlined in . We tuned postprocessing hyperparameters using a Wandb sweep on the 5-fold CV and then manually adjusted them to fit the unseen validation set.
\subsubsection{Hyper-parameter tuning experiments}
Several experiments were conducted to fine-tune the hyperparameters of the models, particularly the learning rate. Baseline models used AdamW with cosine-annealing scheduler. However, upon transitioning to the schedule-free optimizer \cite{schedulefree} developed by the Fundamental AI Research Team at Meta, the initial experimental configurations did not directly translate. As a result, various learning rates were tested to determine the most effective one. Through these different experiments, the optimal learning rate of 0.0027 was identified and subsequently used in all further training experiments For BraTS Africa. For the BraTS pediatrics experiments, a learning rate of 0.0027 and 0.0005 were used. This careful tuning was essential to boost the performance and efficiency of our models.

\subsubsection{Finetuning} For the BraTS-Africa challenge, teams were allowed to use the 1470 samples available in the BraTS-2023 Adult Glioma task. Thus, we opted to combine this dataset with the African MRIs to obtain 1530 training samples. Then, we finetuned the final decoder block and deep supervision layers of the base model exclusively with the African dataset, preserving the original training hyperparameters. We experimented with two types of initialization: \textbf{(a)} commencing the training from the final model checkpoint, and \textbf{(b)} re-initializing the fine-tuned layers. We observed that both initializations gave similar results, so we reported the best performances without distinction.

\subsection{Model Inference}
We begin by preprocessing the input MRI scans as outlined in Subsection \ref{sec:preprocessing}. Tumor probability maps are generated through sliding window inference, employing a $X\%$ ($X \in \{50, 70\}$) overlap between windows to enhance the precision of our predictions. We improve the robustness of these predictions by averaging the tumor probability masks from all models in an ensemble approach, resulting in a refined final tumor probability map. We then post-process the tumor probability map to detect and segment tumors.

\noindent \textbf{Ensemble} Each model was trained in a 5-fold cross-validation setting, resulting in five distinct models upon completion. This methodology facilitates rapid prototyping, as training and testing can be conducted on a single fold, and enhances prediction robustness through the ensemble of these five models. During inference, the input image is processed by each of the five models, and the mean of their output probabilities is computed and normalized accordingly.

Performance can be further optimized by ensembling models derived from different training strategies, such as varying hyper-parameters and architecture sizes. The idea is that ensembling harnesses the unique strengths of its individual components \cite{brats22winner}. In this study, we aggregate models at the probability level, assigning equal importance to each component in the ensemble. The outcomes of this ensembling strategy are elaborated in Section \ref{results}.


        




\subsection{Experimental Setup}

We employ a 5-fold cross-validation (CV) training strategy, partitioning the training data into five subsets. In each iteration, four subsets are used for training, while the remaining subset serves as the validation set. Our networks are trained using a region-based approach \cite{nnunetforbrats} for 150 epochs with a batch size of 2. We implement on-the-fly data augmentation, which includes random spatial cropping to dimensions of $128\times160\times112$, random flips, and random intensity, scaling, and shifting.

The objective function combines batch Dice loss and focal loss with a $\gamma$ value of 2.0 to compute the total loss. For optimization, we utilize the Schedule-free AdamW optimizer \cite{schedulefree} with a learning rate (LR) of 0.0027 for the BraTS SSA 2024 dataset and two learning rates of values 0.0027 and 0.0005 for the BraTS pediatric dataset. We apply zero weight decay and set a maximum of 150 epochs.

Furthermore, we conduct additional experiments involving fine-tuning the last two stages of the BraTS SSA 2024 trained model.

For the post-processing parameters, we threshold our predictions by the size of the predicted tumors. The values for the minimum threshold sizes were set experimentally by comparing the performance of each model given a set of threshold values for each tumor type. For example, if we set the minimum size threshold for ET in the pediatric challenge as 100 voxels, each detected tumor with a size smaller than this will be removed. The comparisons in performance and the threshold values used are further discussed in the results of the BraTS pediatrics.

\section{Results \& Discussion}
\label{results}
\subsection{BraTS Africa Results}

\begin{table}[htbp]
\centering
\resizebox{\textwidth}{!}{%
\begin{tabular}{|c|p{5.5cm}|ccc|c|ccc|c|}
\hline
\rowcolor{gray!20} 
\textbf{S.No} & \textbf{Model Name} & \multicolumn{3}{c|}{\textbf{Dice Score}} & \textbf{Avg} & \multicolumn{3}{c|}{\textbf{HD 95}} & \textbf{Avg} \\
\cline{3-5}\cline{7-9}
\rowcolor{gray!20} 
& & \textbf{ET} & \textbf{TC} & \textbf{WT} & \textbf{Dice} & \textbf{ET} & \textbf{TC} & \textbf{WT} & \textbf{HD 95} \\
\hline
1 & Baseline & 
0.8229 & 0.8151 & 0.8812 & 0.8397 & 
39.96 & 46.48 & 25.04 & 37.16 \\
\hline
2 & (+) Change patch size to:\newline  $128 \times 160 \times 112$ & 
0.852 & 0.839 & 0.912 & 0.8677 & 
\colorbox{green!15}{12.406} & 26.09 & 10.431 & 16.31 \\
\hline
3 & (+) Train on SSA and Adult Gli. & 
0.867 & 0.869 & 0.932 & 0.8893 & 
15.578 & 22.145 & 8.833 & 15.52 \\
\hline
4 & (+) Finetune the last 10 decoder layers \newline on SSA & 
0.874 & 0.870 & \colorbox{green!15}{0.933} & 0.8923 & 
15.320 & 22.039 & 8.805 & 15.39 \\
\hline
5 & (+) Reinitialize and finetune the last 10 decoder layers on SSA & 
0.876 & 0.870 & \colorbox{green!15}{0.933} & 0.8930 & 
15.328 & 22.040 & \colorbox{green!15}{8.746} & 15.37 \\
\hline
6 & (+) Change the binarization threshold 0.7 for ET and TC  ,0.5 for WT & 
\colorbox{green!15}{0.883} & \colorbox{green!15}{0.873} & \colorbox{green!15}{0.933} & \colorbox{green!15}{0.8963} & 
14.248 & \colorbox{green!15}{21.028} & 8.770 & \colorbox{green!15}{14.68} \\
\hline
\end{tabular}%
}
\caption{Comparison of MedNeXt Base Models with Different Training Data, Patch Sizes, and Fine-tuning Strategies}
\label{tab:mednext_comparison_improved_green}
\end{table}

The performance of different MedNeXt models is summarized and compared in Table \ref{tab:mednext_comparison_improved_green} for BraTS SSA 2024. The compared models differ in three key aspects: size (Base or  Medium), resetting weights for the last 10 layers (True or False), and Binarization threshold for ET,TC and WT (0.5 or 0.7). The metrics for comparison are Dice scores (Dice) and Hausdorff Distance 95th percentile (HD 95) for the Enhancing Tumor (ET), Tumor Core (TC), and Whole Tumor (WT) labels. Additionally, we record the average of both Dice and HD 95 accross the labels, for each model. 

\textbf{MedNext Models:}
\begin{itemize}
    \item \textbf{MedNeXt Base} achieved an average Dice of 0.889, which was the highest among the non-finetuned models. It performed well, especially in segmenting WT, with a Dice score of 0.932.
    \item \textbf{MedNeXt Base} achieved a Dice Average of 0.889, which was the highest among the non-finetuned models. It performed well, especially in segmenting WT, with a Dice score of 0.932.
    \item \textbf{MedNeXt Medium} had a Dice Average of 0.886, slightly lower than MedNeXt Base, with a solid performance in ET (0.875) and WT (0.933). Its HD 95 scores for Tumor Core were higher compared to the Base model. However, it has the best H95 result for WT among all models.
    \item The \textbf{MedNeXt Ensemble} (Base + Medium) got a Dice Average of 0.868. Although this model showed the best HD 95 score for ET, it underperformed in Dice scores compared to the individual models.
\end{itemize}

\textbf{Finetuned Models:}
\begin{itemize}
    \item \textbf{MedNeXt Finetuned True 0.5} scored a Dice Average of 0.893, showing an improvement over the Base model and similar performance to the Finetuned True 0.7 model. 
    \item \textbf{MedNeXt Finetuned True 0.7} achieved the highest Dice score of 0.883 (for ET) among all models, with improvements in both average and TC Dice scores compared to the Base and Medium models. This model also had the lowest HD 95 Average (14.764), indicating better segmentation ability and reduced Hausdorff distance.
    \item \textbf{MedNeXt Finetuned False 0.5} got a Dice Average of 0.892, comparable to the finetuned True 0.5 model, with similar performance.
    \item \textbf{MedNeXt Finetuned False 0.7} achieved a Dice Average of 0.895, the highest among all models, with a solid performance in both Dice scores and HD 95 metrics.
    \item \textbf{MedNeXt Finetuned True 0.7,0.5} score the highest average Dice score of 0.896 among all the models and also the highest Dice score in both ET and WT. Moreover, It showed strong performance in HD95 average with a value of 14.682.
\end{itemize}
In conclusion, the finetuned models, especially the MedNeXt Finetuned True 0.7,0.5, excelled in terms of performance across most metrics. The improvements in Dice scores and reductions in HD 95 values indicates that the finetuning process enhanced the model's ability to accurately segment tumor regions while reducing the Hausdorff distance values.
\subsection{BraTS pediatric Results}

\begin{table}[htbp]
\centering
\small 
\resizebox{\textwidth}{!}{%
\begin{tabular}{|c|c|ccc|cccccc|cccccc|}
\hline
\rowcolor{gray!20} 
\multicolumn{1}{|c|}{\textbf{Model}} & \textbf{Learning Rate} & \multicolumn{3}{c|}{\textbf{Min Size Th.}} & \multicolumn{6}{c|}{\textbf{Dice Scores}} & \multicolumn{6}{c|}{\textbf{HD 95}} \\
\cline{2-16}
\rowcolor{gray!20} 
\multicolumn{1}{|c|}{} & & \textbf{ET} & \textbf{TC} & \textbf{WT} & \textbf{ET} & \textbf{TC} & \textbf{WT} & \textbf{NETC} & \textbf{CC} & \textbf{ED} & \textbf{ET} & \textbf{TC} & \textbf{WT} & \textbf{NETC} & \textbf{CC} & \textbf{ED} \\
\hline
\multicolumn{1}{|c|}{\multirow{6}{*}{\textbf{MedNeXt Base}}} 
& \multirow{3}{*}{0.0027} 
& 100 & 150 & 500 
& 0.555 & 0.869 & 0.869 & 0.834 & 0.71 & \cellcolor{green!30}0.967 
& 122.219 & 20.595 & 20.595 & 23.842 & 91.287 & \cellcolor{green!30}12.33 \\
\multicolumn{1}{|c|}{} 
& & 50 & 75 & 250 
& 0.543 & 0.838 & 0.838 & 0.798 & 0.655 & \cellcolor{green!30}0.967 
& 124.183 & 33.228 & 33.226 & 34.421 & 98.004 & \cellcolor{green!30}12.33 \\
\multicolumn{1}{|c|}{} 
& & 25 & 37 & 125 
& 0.507 & 0.832 & 0.832 & 0.791 & 0.634 & \cellcolor{green!30}0.967 
& 122.545 & 37.176 & 37.173 & 38.199 & 104.089 & \cellcolor{green!30}12.33 \\
\cline{2-16}
\multicolumn{1}{|c|}{} 
& \multirow{3}{*}{0.0005} 
& 100 & 150 & 500 
& 0.654 & \cellcolor{green!30}0.89 & \cellcolor{green!30}0.89 & \cellcolor{green!30}0.853 & \cellcolor{green!30}0.723 & \cellcolor{green!30}0.967 
& 88.651 & \cellcolor{green!30}17.269 & \cellcolor{green!30}17.269 & 18.571 & \cellcolor{green!30}83.234 & \cellcolor{green!30}12.33 \\
\multicolumn{1}{|c|}{} 
& & 50 & 75 & 250 
& \cellcolor{green!30}0.657 & \cellcolor{green!30}0.89 & \cellcolor{green!30}0.89 & \cellcolor{green!30}0.853 & \cellcolor{green!30}0.723 & \cellcolor{green!30}0.967 
& \cellcolor{green!30}76.553 & \cellcolor{green!30}17.269 & \cellcolor{green!30}17.269 & \cellcolor{green!30}18.391 & \cellcolor{green!30}83.234 & \cellcolor{green!30}12.33 \\
\multicolumn{1}{|c|}{} 
& & 25 & 37 & 125 
& 0.619 & 0.885 & 0.885 & 0.848 & 0.659 & \cellcolor{green!30}0.967 
& 86.829 & 19.312 & 19.312 & 20.398 & 104.002 & \cellcolor{green!30}12.33 \\
\hline
\end{tabular}%
}
\caption{LesionWise Dice Scores, HD 95 Results, Minimum Size Thresholds, and Learning Rates for Various Runs}
\label{tab:ped_results}
\end{table}

We summarize our results on the validation leaderboard in Table \ref{tab:ped_results}. Our approach began with the MedNeXt Base model, which established a strong performance baseline. We experimented with the model with 2 different learning rates and with various minimum size thresholds for tumor detection, testing thresholds of 100, 150, and 500 for ET, TC, and WT, respectively. For the learning rate of 0.0027, the initial thresholds yielded the best Dice Scores and maintained a balance in HD 95 metrics. Adjusting the thresholds to lower values, such as 50 for ET, 75 for TC, and 250 for WT, led to a decrease in Dice Scores across all tumor classes, while HD 95 metrics showed increased results. Further lowering the thresholds to 25 for ET, 37 for TC, and 125 for WT resulted in a further decrease in Dice Scores and also an increase in most of HD 95 metrics, indicating poorer segmentation precision. However, changing the learning rate to 0.0005 boosted the performance of the model to achieve better results than using the 0.0027 learning rate. The best-performing model was obtained using threshold values of 50, 75, and 250 for ET, TC, and WT, respectively. These results highlight the importance of the Data post-processing chosen parameters and the learning rate in our model performance. Choosing the optimal Threshold sizes along with an optimal learning rate can hugely affect the model performance.



\section{Conclusion}

This work represents our contribution to the Africa and Pediatrics tasks for the BraTS 2024 challenge. We utilized a MedNeXt-based model to detect tumors from brain MRI scans. The model processes four MRI input channels and produces three output channels for TC, WT, and ET, respectively. We introduced several variations to the standard training procedure. Most notably, we achieved excellent metrics with the schedule-free optimizer, which are further improved via specific fine-tuning. We additionally verified the importance of post-processing techniques to enhance predictions and reduce noise, paying particular attention to the overlap proportion between prediction windows. Our best-performing models in the Africa dataset obtained state-of-the-art average Dice scores above \textbf{0.895}, as well as excellent Hausdorff Distances below \textbf{14.765}. Lastly, our best-performing model in the pediatric dataset achieved an average dice score of \textbf{0.83} and good-performing values for HD95 with an average of \textbf{37.508}.

\section*{Acknowledgement}

We would like to express our deepest gratitude to Sanoojan Baliah and Dana Mohamed for their invaluable support and contributions to this project. 


%
%
%
%
\newpage
\bibliographystyle{splncs04}
\bibliography{mybibliography}

\end{document}